\documentclass[prb,twocolumn]{revtex4}
\usepackage{graphicx}
\usepackage{amsmath}
\usepackage{amssymb}

\begin{document}

\title{Rectification in single molecular dimers with strong polaron effect}
\author{Gregers A. Kaat and Karsten Flensberg}
\affiliation{Nano-Science Center, Niels Bohr Institute, Universitetsparken 5, 2100
Copenhagen, Denmark}

\begin{abstract}
We study theoretically the transport properties of a molecular two level
system with large electron-vibron coupling in the Coulomb blockade regime.
We show that when the electron-vibron coupling induces polaron states, the
current-voltage characteristic becomes strongly asymmetric because, in one
current direction, one of the polaron state blocks the current through the
other. This situation occurs when the coupling between the polaron states is
smaller than the coupling to the leads. We discuss the relevance of our
calculation for experiments on $C_{140}$ molecules.
\end{abstract}

\maketitle

\section{ Introduction}

The possibility of designing molecular junctions that behave as rectifiers
was suggested by Aviram and Ratner\cite{aviramratner74} based on an
asymmetric two-level donor molecular system, with one level empty and one
filled (an acceptor and a donor level). In one bias direction the levels are
tuned closer to resonance, whereas for the reverse bias they are further
detuned, giving rise to an asymmetric current-voltage (IV) characteristic.
This scenario has been investigated in molecular systems for a number of
years.\cite{metz03} The transport properties depend crucially on both the
Coulomb interactions and the relaxation of the nuclei during the electron
transfer processes. The Coulomb interactions limit the allowed charge
configurations, while the electron-vibron coupling determines the overlap
between the different configurations as well as provides channels for energy
relaxation.

Molecular transistors with strong electron-vibron coupling has
recently become an active research area, both experimentally
\cite{park00,kuba03,yunatt,pasu03} and
theoretically.\cite{wing89,boese01,lund02,flen03,braig03,mcca03,alex03,mitra04,galp04,ness01,wege}
Different types of vibrational modes have been observed. In the
original work of Park et al.\cite{park00} clear signatures of
phonon sidebands were seen in the tunnel spectrum. The size of the
individual steps in the $IV$ curves agreed well with a simple
model based on Franck-Condon physics. \cite{park00} Furthermore,
sidebands caused by internal vibration of the molecular systems
have also been identified experimentally and, in particular, in a
recent experiment by Pasupathy et al.\cite{pasu03} where tunneling
through dimerized $C_{70}$ molecules was measured. These authors
claimed that a vibrational mode at 11 meV due to the relative
motion of the two $C_{70}$ molecules was seen in the experiment.

In this paper, we study theoretically transport through a dimer
molecule with a single internal vibration. As a generic model, we
include a coupling between the charge on the two parts of the
dimer. The electron-vibron coupling thus introduces a possibility
of forming polaronic states. Transport in systems with
polarons\cite{holstein} is a well-studied subject and the polaron
formation is known to reduce the mobility. In single-electron
transistor systems polaronic effects have also been studied
theoretically, in particular for single-level systems
\cite{wing89,boese01,lund02,flen03,braig03,mcca03,alex03,mitra04,galp04},
and only recently for a two level system.\cite{wege} Here we study
a similar system but with the important modification that the two
levels in addition are coupled by tunneling, in which case an
internal polaron can form. The main point of our study is that
when this electron-vibron coupling is sufficiently strong and a
polaron state is formed, it may lead to rectification effects and
negative differential conductance, depending on how the molecule
is situated in the constriction.

The paper is organized as follows. In Section II we set up the model
describing the dimer and the electron-vibron coupling, in Section III we
discuss the physical concepts in a semi-classical picture, in Section IV we
calculate the transport in the situation where the tunneling rates is
smaller than the polarons couplings, whereas Section V discusses the
situation when the polarons are localized on the time scale set by tunneling
to the leads. The cross-over between these two regime is studied in Section
VI and finally Section VII discusses the experimental relevance and
concludes.

\section{The molecular dimer model}

\begin{figure}[tbp]
\centerline{\includegraphics[width=.45\textwidth,clip=true]{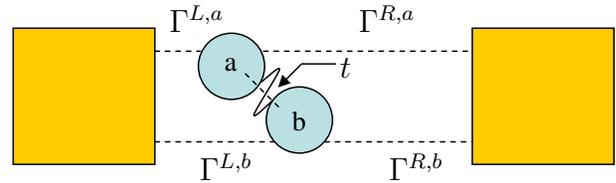}}
\caption{Dimer model studied in this paper. The two parts of the dimer are
represented by single electronic levels, $a$ and $b$, connected by a
tunneling amplitude, $t$. The intramolecular vibration is illustrated as a
spring connecting the two levels. A force acts on the part of the dimer that
is occupied by an electron due to an electric field in the gap. }
\label{fig:dimermodel}
\end{figure}
The molecular model that we study (see Fig.~\ref{fig:dimermodel}) is a dimer
with 2 electron sites, or in a different terminology a ``double quantum
dot". For simplicity, we neglect both spin and possible orbital degeneracies
in the model. The two sites are coupled by tunneling. Furthermore, due to an
electric field across the gap the internal molecular vibrational mode
couples to the charges difference on the two sites. The electric field is
produced either by the bias voltage, local charge traps, or image charges
created in the leads by the charged molecule. In total our model Hamiltonian
for the molecule reads
\begin{align}
H_D^{{}}&= \frac{p^2}{2m} +\frac12 m\omega^2x^2+
\Delta(n_a^{{}}-n_b^{{}})+eV_g(n_a^{{}}+n_b^{{}})  \notag \\
&\quad +Un_a^{{}} n_b^{{}}- t (c_a^{\dagger} c_b^{{}} +c_b^{\dagger}
c_a^{{}}) +\lambda(n_a^{{}}-n_b^{{}})x,
\end{align}
Here $c_j^\dagger$ ($c_j^{{}}$) ($j$=$a,b$) are the electron creation
(annihilation) operators, whereas $b^\dagger(b)$ is the vibron
creation(annihilation) operator and $n_j^{{}} = c_j^\dagger c_j^{{}}$.
Furthermore, $V_g$ is the gate voltage which couples to the total charge on
the molecule, $\Delta$ describes the energy difference between the two sites
and, finally, $U$ the Coulomb repulsion between electrons occupying the two
levels. The total Hamiltonian describing the molecule, the leads and the
coupling between them is
\begin{equation}  \label{H}
H=H_D^{{}}+\sum_{\eta=L,R}\left(H_{\eta}+H_{T\eta}\right),
\end{equation}
where
\begin{align}
H_{\eta}^{{}} &= \sum_{k} \xi_{k\eta}^{{}}
c^\dagger_{k\eta}c^{{}}_{k\eta},\quad \eta=L,R, \\
H_{T\eta}^{{}}&=\sum_{k,j=a,b} \left(
T_{k\eta,j}^{{}}\,c^\dagger_{k\eta}c_j^{{}}+T_{k\eta,j}^*\,c^\dagger_{j}c_{k
\eta}^{{}} \right).
\end{align}

Throughout, we will assume large $U$ and the gate voltage tuned such that
the total occupancy is either zero or one. The subspace with zero electrons
is, of course, easily diagonalized. With one electron on the dimer, the
situation is more complicated. Introducing Pauli matrices for the electronic
degree of freedom and the dimensionless parameters (using $\hbar =1$)
\begin{equation}
\alpha =\frac{\lambda ^{2}}{m\omega ^{2}t},\quad g=\frac{\lambda
^{2}}{2m\omega ^{3}},\quad \delta =\frac{\Delta }{\omega },
\label{param}
\end{equation}
we are left with the Hamiltonian
\begin{equation}
H_{D}^{{}}/\omega =b^{\dagger }b+\frac{1}{2}+\sqrt{g}\,
\hat{\sigma}_{z}(b^{\dagger }+b)-\frac{2g}{\alpha
}\,\hat{\sigma}_{x}+\delta \,\hat{\sigma}_{z} +eV_g,
\label{Hmodel}
\end{equation}
where $b$ and $b^\dagger$ are the usual boson annihilation and
creation operators. This Hamiltonian can be solved numerically in
a truncated boson Hilbert space, which is done below in order to
determine the eigenenergies and the overlap factors between the
empty and the occupied molecule. At $\alpha =1$ there is a
cross-over to a regime where the fermion and boson degrees of
freedom become strongly correlated due to the formation of polaron
states. When the levels are non-degenerate the polaron leads to an
increased localization of the electronic wavefunction. In both
cases there is a strong reduction of the effective tunneling
amplitude between the two sites, which has consequences for
conductance through the molecule.

\section{Semi-classical analysis}

In order to understand the polaron formation, we start by a
semiclassical analysis of the spin-boson model. This is done in
order to illustrate the physics and will not be used in the actual
description described in the following sections. The semiclassical
treatment is formally valid when $\omega\ll t_\mathrm{eff}$, but
in the examples taken below this is not the case, and we instead
resort to numerical diagonalization of the Hamiltonian in
Eq.~\eqref{Hmodel}.

\begin{figure}[tbp]
\centerline{\includegraphics[width=.5\textwidth]{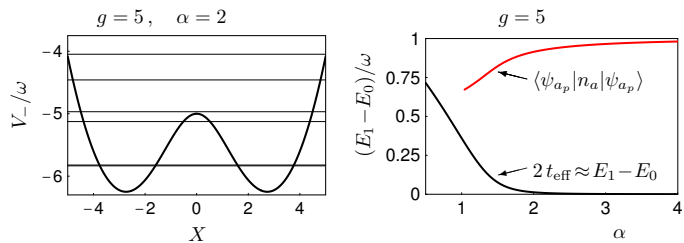}}
\caption{Left panel: The semiclassical potential for $g=5$ and
$\protect\alpha =2$ and the exact eigenvalues (horizontal lines)
found by numerical diagonalization of Eq.~\eqref{Hmodel}. Note
that the two lowest eigenvalues are almost degenerate,
corresponding to two polaron states split by a small effective
tunneling coupling $t_{\mathrm{eff}}$. The effective coupling is
shown in the right panel as a function of $\protect\alpha $
together with the electron population of level $a$ in the polaron
state $\protect\psi_{a_{p}}$, defined in
Eq.~\eqref{polaronstates}. The remaining eigenenergies are not
degenerate and therefore the corresponding eigenstates are
delocalized on the molecule. } \label{fig:semifig}
\end{figure}

Treating the harmonic oscillator classically, the model can be solved in the
electronic Hilbert-space. The electronic Hamiltonian then becomes,
\begin{align}
H_{De}= \omega \left(
\begin{array}{cc}
\sqrt{2g}\, X + \delta & -2g/ \alpha \\
-2g/\alpha & -\sqrt{2g}\, X - \delta \\
&
\end{array}
\right),
\end{align}
where $X=x/\ell$ and $\ell=\sqrt{1/m\omega}$. For each of the two electronic
eigenstates we thus have an effective potential for the oscillator degree of
freedom. These Born-Oppenheimer surfaces are
\begin{equation}
\frac{V_\pm}{\omega} = \frac12 {X^2} \pm \sqrt{2 g {X^2}+\frac{4g^2}{\alpha
^2}+2 \sqrt{2g} X \delta +{\delta^2}}.
\end{equation}
In the electronic groundstate $V_-$ there is only one minimum for
$\alpha<1$ , whereas for $\alpha>1$ there are two minima. Thus a
bifurcation occurs at $\alpha=1$. In the bifurcated domain, the
electronic and bosonic degrees of freedom become highly
correlated, because when the oscillator is localized in one of the
two minima, the electron wavefunction is changed accordingly. Thus
the physics is similar to that of a small polarons (see
Fig.~\ref{fig:semifig}). In Fig.~\ref{fig:semifig} we show the
bifurcated potential, the exact numerical eigenvalues, and the
effective tunneling coupling between the polaron states. From our
exact diagonalization we thus find that the difference in energy
between the groundstate and the first excited state becomes
exponentially small in the polaron regime.

In the semi-classical picture, the splitting of the polaron states
occurs because of tunneling. The tunneling amplitude is small,
because in order to move, it must drag the oscillator displacement
with it. Furthermore, a coupling to a dissipative environment will
tend to localize the polaron even more, because it couples to
displacement coordinate and thus tends to destroy the coherence
between the two states,\cite{leggettwolevel} resulting in a small
effective tunneling coupling $t_{\mathrm{eff}}$. When the
effective coupling becomes very small, the two lowest eigenstates
$|\psi_0\rangle$ and $|\psi_1\rangle$ form two polaron states
which are approximate eigenstates:
\begin{equation}  \label{polaronstates}
|\psi_{a_p(b_p)}\rangle=( |\psi_0\rangle\pm |\psi_1\rangle)/\sqrt{2}.
\end{equation}

As long as the splitting in energy is larger than the tunneling broadening
of the levels, the polaron formation is not important for the transport
other than it influences the Franck-Condon factors and the molecule can
still be regarded as one quantum system (see Fig.~\ref{fig:single}). This
situation is analyzed in Section \ref{sec:oneq}. However, if the effective
tunneling rate $t_{\mathrm{eff}}$ is smaller than the tunneling rates to and
from the leads, a different physical picture emerges, because the dimer will
behave as a "double dot" with weak interdot coupling but strong Coulomb
interaction (see Fig.~\ref{fig:double}). Thus a master equation treatment
should take into account that the molecular system has more internal states,
which is done in Section \ref{sec:twoq}. This regime is where the
rectification occurs as discussed in detail below.

\section{Transport properties for $\Gamma \ll t_{\mathrm{eff}}$}

\label{sec:oneq}

As argued, the transport properties of the dimer transistor depend on the
ratio between tunneling rates and internal time scales of the molecule. Let
us first assume that all tunneling rates $\Gamma^{R a}, \Gamma^{R b},
\Gamma^{L a}$ and $\Gamma^{L b}$ are much smaller than $t_\mathrm{eff}$.
Furthermore, throughout we consider the weak tunneling limit in the sense
that the vibrational degrees of freedom are assumed to relax between
tunneling events, i.e. $\omega/Q \gg\Gamma$, where $Q$ is the quality factor
of the vibrational mode.\cite{braig03} If this limit is not satisfied the
non-equilibrium vibron distribution should also be determined, however, for
the physics discussed here this is not important. Also, the broadening of
the vibron sidebands is not included and assumed dominated by the thermal
smearing such that $k_\mathrm{B}T\gg \omega/Q$. To describe this situation,
we now use the master equation approach and only the occupations of the
different charge states need to be determined by the kinetics. Allowing only
two charge states, with occupations $P_0$ and $P_1$, the master equation is
\begin{align}
-P_{1}^{{}}\Gamma^{}_{01} - P_{0}^{{}}\Gamma_{10}=0,  \label{master0}
\end{align}
together with the normalization condition
\begin{equation}  \label{normalization0}
P_0+P_1{}=1.
\end{equation}
Here $\Gamma_{ij}$ is the tunneling rate from state $j$ to state $i$. The
expression for the current then becomes
\begin{equation}  \label{Iseq}
I = (-e) \frac{\Gamma_{10}^R \Gamma_{01}^L - \Gamma_{01}^R
\Gamma_{10}^L}{\Gamma_{10}^R + \Gamma_{10}^L + \Gamma_{01}^R +
\Gamma_{01}^L},\quad (\Gamma\ll t_\mathrm{eff}).
\end{equation}
In Eqs.~(\ref{master0}),(\ref{Iseq}) the tunneling rates are
calculated using Fermi's golden rule. When calculating the rates
in Eq.~\eqref{Iseq} we encounter the tunneling densities of states
of the form: $\Gamma^\eta=2\pi\sum_{kj}
|T_{k\eta,j}|^2\delta(\xi_{k\eta})$, ($\eta=L,R$). The evaluation
of the cross terms $\sum_k T_{k\eta,a}^*T_{k\eta,b}^{{}}
\delta(\xi_{k\eta})$ requires detailed knowledge of the relative
phases of the tunneling amplitudes to the two parts of the dimer.
Fortunately, a simplification is possible because
$T_{k\eta,a}^*T_{k\eta,b}^{{}}\sim \exp(ik d)$, where $d$ is the
distance between the two parts of the dimer and $k$ is the wave
number of the lead electrons. Typically, one has $kd\gg 1 $ and
the cross-terms average to zero when summing over $k$. Thus, no
interference effects between the two tunneling paths onto the
molecule is expected. With these observations, we obtain
\begin{equation}
\Gamma_{10}^{\eta} = \sum_{j=a,b}\Gamma^{\eta j} \sum_{i_0^{{}},f_1^{{}}}
\left|\langle f_1^{{}}|c_j^{\dagger}|i_0^{{}}\rangle \right|^2 \frac{e^{-
\beta E_{i_0}^{{}} }}{Z_0}\, n_\eta^{{}} (E_{f_1}^{{}} - E_{i_0}^{{}}),
\end{equation}
and
\begin{align}
\Gamma_{01}^{\eta} &= \sum_{j=a,b}\Gamma^{\eta j} \sum_{i_1^{{}},f_0^{{}}}
\left|\langle f_0^{{}}|c_j^{{}}|i_1^{{}}\rangle\right|^2\frac{e^{- \beta
E_{i_1}^{{}}}}{Z_1}  \notag \\
&\qquad \times\left[ 1- n_\eta^{{}}( E_{i_1}^{{}}- E_{f_0}^{{}})\right].
\end{align}
Here, $|i_0 \rangle$, $|i_1 \rangle$ ($|f_0 \rangle$, $|f_1 \rangle$) denote
initial (final) states of the empty (0) and filled (1) dimer, and
\begin{equation}  \label{neta}
n_\eta(E)=1/(\exp((E-eV_\eta)/k_\mathrm{B}T)+1)
\end{equation}
are the Fermi functions in the lead $\eta$ with applied voltages $V_\eta$.
Finally, $E_{i_n}$ and $E_{f_n}$ are the total dimer energies for the
initial and final states with $n$ electrons and $Z_0$ and $Z_1$ the
corresponding partition functions.
\begin{figure}[tbp]
\centerline{\includegraphics[width=.5\textwidth]{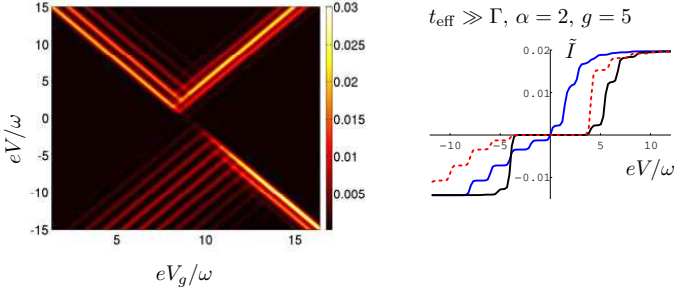}}
\caption{(Color online) Left panel: Contour plot showing the
differential conductance for the dimer molecule in the regime
Eq.~\eqref{Iseq} where holds. The parameters are $g=5$,
$\protect\alpha =2$, $\Delta =0$, $\Gamma ^{Ra}=0$, $\Gamma
^{Rb}/\Gamma ^{La}=0.04$, $\Gamma ^{Lb}/\Gamma ^{La}=0.96$, and
$T=0.1\protect\omega $. The bias voltage is applied symmetrically,
so that $V_L=-V_R=V/2$. Notice the suppression of the differential
conductance at low bias voltage which is due to a overlap between
the filled and empty states. The right panel shows $IV$ curves for
the same parameters (but $T=0.05\omega $) for $V_{g}-V_{g0}=0$
(curve without gap at $V=0 $), $-2\omega $ (dashed) and
$2\protect\omega $. Here $V_{g0}$ defines the gate voltage at
which the groundstates of the two charge states are degenerate.}
\label{fig:single}
\end{figure}

Because we are interested in the asymmetric situation, we study
the situation where one electrode couples stronger than the other
to the molecule, $\Gamma ^{R}\ll \Gamma ^{L}$. Furthermore, we
also allow for a skewed configuration as in
Fig.~\ref{fig:dimermodel}, and in this geometry, we have $\Gamma
^{Ra}<\Gamma ^{Rb}$ and $\Gamma ^{La}>\Gamma ^{Lb}$. In
Fig.~\ref{fig:single}, we show examples of the differential
conductance in the $V-V_{g}$ plane in the situation where the
intramolecular coupling between polaron states is strong, i.e.
when $\Gamma \ll t_{\mathrm{eff}}$ and Eq.~(\ref{Iseq}) applies.
Furthermore, the bias voltage is applied symmetrically, so that
$V_L=-V_R=V/2$.

\section{Transport properties for $\Gamma \gg t_{\mathrm{eff}}$}

\label{sec:twoq}

In the opposite case $\Gamma\gg t_\mathrm{eff}$, the situation is
very different. In this case, an electron that tunnels onto level
$a$ does not resolve the tunneling splitting of the two polaron
states and the molecule no longer relaxes between tunneling and
therefore the two polaron states can be considered as decoupled.
In this case, we must treat the filled molecule as a quantum
system having \textit{more} possible states, since it can be
occupied in one of the two polaron states, from now on denoted by
$a_p$ and $b_p$. Furthermore, the molecule may be occupied in one
of the delocalized states (see Fig.~\ref{fig:semifig}).

Moreover, we consider the following hierarchy of energies $\omega>
k_\mathrm{B}T> \omega/Q> \Gamma> t_\mathrm{eff}$. The last two
inequalities imply that between tunnelings the molecule relaxes to
a thermal distribution within each polaron subspace. Since the
current drives the occupancies of the two polaron states out of
equilibrium, we need to determine the distribution functions of
the two polaron states, $P_{a_p}^{{}}$ and $P_{b_p}^{{}}$,
respectively. If the molecule is occupied in a delocalized state
it is in either polaron state with equal probability. The new set
of master equations thus reads (with $\tau=a_p$ or $\tau=b_p$)
\begin{subequations}
\label{doubledotmaster}
\begin{align}
-P_{\tau}^{{}}\Gamma^{\tau}_{01} - P_{0}^{{}}\Gamma^{\tau}_{10}&=0, \\
-P_{0}^{{}}\sum_{\tau}\Gamma^{\tau}_{01} - \sum_{\tau}
P_{\tau}^{{}}\Gamma^{\tau}_{10}&=0,
\end{align}
together with the condition
\begin{equation}  \label{normalization}
P_0+\sum_{\tau} P_{\tau}^{{}}=1.
\end{equation}
The tunneling rates in Eq.~\eqref{doubledotmaster} have contributions from
both leads:
\begin{equation}  \label{sumG}
\Gamma^\tau=\sum_{\eta=L,R}\Gamma^{\eta,\tau}.
\end{equation}
The tunneling rate for tunneling from lead $\eta$ into $a_p$, is (again the
oscillating cross-terms are ignored)
\end{subequations}
\begin{equation}
\Gamma_{10}^{\eta,a_p} = \sum_{j=a,b}\Gamma^{\eta j}\!\!
\sum_{i_0^{{}},f_1^{a_p}} \left|\langle
f_1^{a_p}|c_j^{\dagger}|i_0^{{}}\rangle \right|^2 \frac{e^{- \beta
E_{i_0}^{{}}}}{Z_0}\, n_\eta^{{}} (E_{f_1^{a_p}} - E_{i_0}^{{}}),
\end{equation}
and similarly for tunneling into the polaron states $b_p$. Here
$|f_1^{a_p}\rangle$ and $|f_1^{b_p}\rangle$ mean the polaron
states in Eq.~\eqref{polaronstates} if they are degenerate, i.e.
$(|f_1^{(1)}\rangle\pm |f_1^{(2)}\rangle/\sqrt{2}$, where
$|f_1^{(i)}\rangle$, $i=1,2$ are the two degenerate states
(meaning that eigenenergies differ by less than $\Gamma$). For the
non-degenerate states,
$|f_1^{a_p}\rangle=|f_1^{b_p}\rangle=|f_1\rangle$ implying that
the system ends up in either polaron state with equal probability.

\begin{figure}[t]
\centerline{\includegraphics[width=.5\textwidth]{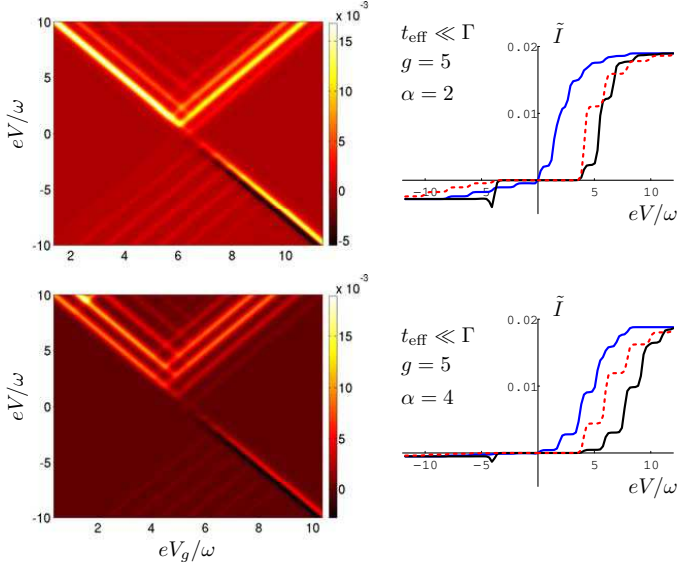}}
\caption{(Color online) Left panel: Contour plots showing the
differential conductance for the dimer molecule when
Eq.~(\protect\ref{doubledotcurrent}) holds, for $\protect\alpha=2$
(top) and $\protect\alpha=4$ (bottom). Other parameters are $g =
5$, $\Delta=0$, $\Gamma^{Ra}= 0$, $\Gamma^{Rb}/\Gamma^{La} =
0.04$, $\Gamma^{Lb}/\Gamma^{La} = 0.96$, and
$T=0.1\protect\omega$. The bias voltage is applied symmetrically,
so that $V_L=-V_R=V/2$. Right panels show corresponding $IV$
curves for different gate voltages as in
Fig.~\protect\ref{fig:single}, and with $T=0.05\protect\omega$.
Note the strong blocking of the current in one direction and the
negative differential resistances. For $\protect\alpha=2$ the
current in the blocked direction is larger than for
$\protect\alpha=4$, because the polaron is less localized on one
electron level (see Fig.~\protect\ref{fig:semifig}).}
\label{fig:double}
\end{figure}

For the tunneling-out processes the rates are
\begin{align}
\Gamma_{01}^{\eta,a_p} &= \sum_{j=a,b}\Gamma^{\eta j}
\sum_{i_1^{a_p},f_0^{{}}} |\langle f_0^{{}}|c_j^{{}}|i_1^{{a_p}}\rangle |^2
\frac{e^{- \beta E_{i_1}^{a_p}}}{Z_1^{a_p}}  \notag \\
&\qquad \times[1-n_\eta^{{}} (E_{i_1^{a_p}}^{{}}-E_{f_0^{{}}})],
\end{align}
where the initial polaron states are defined in the same way as above.
Similar expression holds for tunneling out of the polaron state $b_p$. Thus,
treating the dimer as separate polaron states, we find a different
expression for the current:
\begin{equation}  \label{doubledotcurrent}
I = (-e)\frac{\sum_\tau \left[\Gamma_{10}^{R\tau}
\Gamma_{01}^{L\tau}-\Gamma_{10}^{L\tau} \Gamma_{01}^{R\tau}\right]
/\Gamma^\tau_{01}}{1 + \sum_\tau \left(
\Gamma_{10}^{\tau}/\Gamma_{01}^{\tau}\right)}\quad (\Gamma\gg
t_\mathrm{eff}).
\end{equation}
In Fig.~\ref{fig:double} we show the results in the case where
$\Gamma\gg t_\mathrm{eff}$ and Eq.~\eqref{doubledotcurrent} holds.
The $IV$ characteristics are seen to be strongly asymmetric even
though the bias is applies in a symmetric way, i.e.$V_L=-V_R=V/2$.
The reason for the asymmetry is that the polaron state, which is
weakly coupled to the right lead can block current through the
other, more strongly coupled state. However, this blockade only
happens in one current direction, namely when the electrons run
toward the weakly coupled electrode.

\section{Cross-over regime: Generalized master equation}

\label{sec:GME}

In the regime where $\Gamma $ and $t_{\mathrm{eff}}$ are of the
same order, we cannot use any of the two approaches above, because
when two states in the molecular system are degenerate or closer
than $\Gamma $ in energy, the current through the system has to be
described as one coherent process. This is in general a difficult
problem. However, in the limit of large bias voltage a generalized
master equations\cite{gurv96,stoof,gurv98,wege98} has been
developed. To study how the asymmetry of the large bias
conductance, we here adopt this method. The large bias assumption
implies that the voltage must be large compared to the temperature
and the (effective) tunneling couplings as well molecular
energies. The assumption of large bias leads to the a set of
Markovian master equations for the density
matrix.\cite{gurv96,stoof,gurv98,wege98} Thus at large bias there
is no correlation between individual tunneling events, but still
the transfer of electrons is described coherently. The generalized
master equation approach can, in fact, describe the cross-over we
discuss here, because the blocking effect found in Section V does
not depend on time-correlations between tunnel events, but on how
strongly a single electron is localized on the molecule during
tunneling.

Furthermore, in order to simplify the problem we consider only the two
lowest states which correspond to the two polaron states located to the left
or to the right with some effective coupling $t_{\mathrm{eff}}$. The
generalized master equations are then a modification of those developed in
Ref.~\onlinecite{stoof} for a double dot system, because here we allow for
in- and out tunnelings on both dots. For the case when the electron current
is from left to right (i.e. $V<0$), we have
\begin{subequations}
\begin{align}
\partial _{t}\rho _{aa}& =\Gamma ^{La_{p}}\rho _{0}-\Gamma ^{Ra_{p}}\rho
_{aa}+it_{\mathrm{eff}}(\rho _{ba}-\rho _{ab}), \\
\partial _{t}\rho _{bb}& =\Gamma ^{Lb_{p}}\rho _{0}-\Gamma ^{Rb_{p}}\rho
_{bb}-it_{\mathrm{eff}}(\rho _{ba}-\rho _{ab}), \\
\partial _{t}\rho _{ab}& =(-\frac{1}{2}(\Gamma ^{Ra_{p}}+\Gamma
^{Rb_{p}})+i\Delta )\rho _{ab}+it_{\mathrm{eff}}(\rho _{bb}-\rho _{aa}), \\
\partial _{t}\rho _{ba}& =(-\frac{1}{2}(\Gamma ^{Ra_{p}}+\Gamma
^{Rb_{p}})-i\Delta )\rho _{ba}-it_{\mathrm{eff}}(\rho _{bb}-\rho _{aa}).
\end{align}
The stationary solution for $\rho $ is found by setting $\partial _{t}\rho
_{ij}=0$ and the current is then $I=(-e)(\rho _{aa}\Gamma ^{Ra_{p}}+\rho
_{bb}\Gamma ^{Rb_{p}})$, which becomes for $V<0$
\end{subequations}
\begin{equation}
I_{V<0}=\frac{(-e)\Gamma ^{L}\Gamma ^{R}(\chi _{R}^{{}}{{{\Gamma
^{R}}}}+{\Gamma ^{Ra_{p}}}{\Gamma ^{Rb_{p}}}/\Gamma
^{R})}{({\Gamma ^{R}}+2\Gamma ^{L})\chi _{R}^{{}}\Gamma
^{R}+{\Gamma ^{Ra_{p}}}{\Gamma ^{Rb_{p}}}(1+\frac{\Gamma
^{La_{p}}}{\Gamma ^{Ra_{p}}}+\frac{\Gamma ^{Lb_{p}}}{\Gamma
^{Rb_{p}}})},  \label{coherentcurrent}
\end{equation}
where
\begin{equation}
\Gamma ^{R}=\Gamma ^{Ra_{p}}+\Gamma ^{Rb_{p}},\quad \Gamma ^{L}=\Gamma
^{La_{p}}+\Gamma ^{Lb_{p}},
\end{equation}
and where
\begin{equation}
\chi _{R}^{{}}=\frac{4{t_{\mathrm{eff}}}^{2}}{4\Delta
^{2}+({\Gamma ^{R}})^{2}},  \label{chidef}
\end{equation}
contains all the $t_{\mathrm{eff}}$ and $\Delta $ dependencies. When $\chi
_{R}^{{}}$ is large, the current reduces to the incoherent result of a dot
with a doubly degenerate level:
\begin{equation}
I_{V<0}=\frac{(-e)\Gamma ^{L}\Gamma ^{R}}{{\Gamma ^{R}}+2\Gamma ^{L}}
\end{equation}
In the opposite limit of small $\chi _{R}^{{}}$ the time an electron spends
on the dimer is too short for hopping between the sites to occur, and the
current reduces to
\begin{equation}
I_{V<0}=\frac{(-e)\Gamma ^{L}}{1+\frac{\Gamma ^{La_{p}}}{\Gamma
^{Ra_{p}}}+\frac{\Gamma ^{Lb_{p}}}{\Gamma ^{Rb_{p}}}}
\end{equation}
which is also the large $V$ limit of Eq.~\eqref{doubledotcurrent} or the
sequential tunneling limit of a double dot system in absence of interdot
tunneling.

The current for positive voltage is easily found from
Eq.~\eqref{coherentcurrent} by interchanging left and right,
$L\leftrightarrow R$.

Let us now focus on the limit considered in the examples in the
previous two sections, namely
$(\Gamma^R,t_\mathrm{eff})\ll\Gamma^L$ and
$(\Gamma^{Ra},t_\mathrm{eff})\ll\Gamma^{Rb}$, and in these limits
we find the following ratio between the currents for the two bias
polarizations
\begin{equation}  \label{Iratio}
\left|\frac{I_{V<0}}{I_{V>0}}\right|=
\frac{\frac{\Gamma^{Ra_p}}{\Gamma^R}+\chi_R^{{}}}{\frac{\Gamma^{La_p}}{\Gamma^L}+2\chi_R^{{}}}.
\end{equation}
From this expression it is evident that degree of rectification is
limited by the smallest tunneling-out rate through the right
junction, $\Gamma^{Ra_p} $ or the the inter-polaron tunneling
coupling, $t_\mathrm{eff}$, through the parameter $\chi_R^{{}}$ in
Eq.~\eqref{chidef}. Eq.~\eqref{Iratio} also shows that when
asymmetry is limited by $\chi_R^{{}}$, i.e. when
$\Gamma^{Ra_p}/\Gamma^R\ll\chi_R^{{}}$, increasing $\Delta$ in
fact enhances the rectification, because a finite $\Delta$
suppresses tunneling between the two sites.

\section{Discussion and summary}

Lastly, we discuss the relevance for the experimental system in
Ref.~\onlinecite{pasu03}, where a single-electron transistor setup
was made with single $C_{140}$ molecules as the active element. Of
course, our two level model cannot fully describe the real
molecule, nevertheless a comparison is interesting. Signatures of
a 11 meV intramolecular vibrational mode were observed in
Ref.~\onlinecite{pasu03} and the coupling parameter $g$ was
estimated to be between 1 and 5. \cite{gpolaron} If the polaron
physics discussed here should have relevance for this or similar
dimer systems, we must have $\alpha \gtrsim 2$, or correspondingly
$t\sim $ 50 meV, which is not unrealistic. In fact, rectification
effects very similar to those predicted in the present paper have
indeed been seen in some $C_{140}$ devices.\cite{pasu:private}

In conclusion, we have studied molecular system with an internal vibrational
mode coupled to the charge difference between two hybridized molecular
levels giving rise to polaron formation. The polaron has distinct
consequences for the transport properties, which we have calculated using
different master equation approaches. First, we analyzed the case when the
internal hybridization between polarons is larger than tunneling rates to
the leads. This gives more or less symmetric $IV$ characteristics, but with
a number of vibron sidebands. Secondly, we studied the case when the two
polarons are weakly coupled and the tunneling rates to the leads dominate
the kinetics. Is this case, another set of master equation was needed and
highly asymmetric $IV$ characteristics are predicted. Finally, we have
analyzed the cross-over regime by a generalized master equation capable of
describing the coherent tunneling, but only for large bias voltages.

The rectification mechanism suggested in this paper should be experimentally
observable. Generally, the current-voltage characteristics of complex
molecular transistors with strong coupling between charge and vibrational
degrees of freedom is a promising tool for studying the details of such
devices.

\section{Acknowledgement}

We thank P. McEuen, J. Park, and A. Pasupathy for discussions on the
experiments and T. Novotny and M. Wegewijs for discussions and comments on
the manuscript. The work was supported by the Danish Natural Science
Foundation and the European Commission through project FP6-003673 CANEL of
the IST Priority.\cite{EUjunk} 


\begin{thebibliography}{99}
\bibitem{gpolaron} In extracting $g$, one should, however, be aware that the
peak heigths distribution is not Poissonian.
\bibitem{EUjunk} The views expressed in this publication are those of the
authors and do not necessarily reflect the European Commission's view on the
subject.
\end{thebibliography}
\end{document}